\documentclass[a4paper,10pt,twoside]{cpc-hepnp}

\usepackage{multicol}
\usepackage{graphicx}
\usepackage{booktabs}
\usepackage{amssymb,bm,mathrsfs,bbm,amscd}
\usepackage[tbtags]{amsmath}
\usepackage{lastpage}
\usepackage{CJK}
\usepackage{footmisc}
\newcommand{\est}{t_\mathrm{EST}}

\begin{document}
%\begin{CJK*}{GBK}{song}

\fancyhead[co]{\footnotesize Yinghui Guan~ et al: Study on efficiency of event start time determination at BESIII}

\footnotetext[0] {Received \today}

\title{Study on efficiency of event start time determination at BESIII \thanks{
Supported by Ministry of Science and Technology of China(2009CB825200), Joint Funds of National Natural Science Foundation of China(11079008), Natural Science Foundation of China(11275266) and SRF for ROCS of SEM}}

\author{%
      GUAN Yinghui$^{1}$
\quad LU Xiao-Rui$^{1;1)}$\email{xiaorui@ucas.ac.cn}\\%
\quad ZHENG Yangheng$^{1;2)}$\email{zhengyh@ucas.ac.cn} %
\quad WANG Yi-Fang$^{2}$
}
\maketitle

\address{%
$^1$ University of Chinese Academy of Sciences, Beijing 100049, China\\
$^2$ Institute of High Energy Physics, Chinese Academy of Sciences, Beijing 100049, China\\
}

\begin{abstract}
A method to estimate efficiency of event start time determination at BESIII is developed. This method estimates the efficiency at the event level by combining the efficiencies of various tracks ($e$, $\mu$, $\pi$, K, $p$, $\gamma$) in a Bayesian way. Efficiencies results and difference between data and MC at the track level are presented in this paper. For a given physics channel, event start time efficiency and systematic error can be estimated following this method.
\end{abstract}



\begin{multicols}{2}

\renewcommand{\thefootnote}{\arabic{footnote}}
\section{Introduction}
The Beijing Spectrometer III (BESIII)~\cite{lab1} is a general detector
at the Beijing Electron-Positron Collider II (BEPCII)~\cite{lab2},
running in the $\tau$-charm energy region. BEPCII is a double storage ring collider which operates in
multi-bunch collision mode. The BESIII detector consists of the Main Drift Chamber (MDC),
Time-Of-Flight (TOF) counter, Electromagnetic Calorimeter (EMC) and Muon Chamber (MUC).

In the BESIII data acquisition system, the logic
of the trigger system and time measurement system is such that
the TDC time of a hit signal in the detection apparatus is taken as
the time interval from the trigger start time to the arrival
time of the hit~\cite{lab3}. This trigger start time may differ
from the real collision time.
The event start time (EST) determination algorithm, therefore, has been developed to calculate
the common start time of the recorded tracks in an event (denoted as $\est$).
The basic idea is a backtrace of the measured TDC
information of the hits, in the MDC or TOF, to the time when the track
was produced near the collision point, using the
reconstructed trajectory obtained from the fast tracking
(FST)~\cite{lab4}. More details can be found in Ref.~\cite{lab5,lab6,lab7}.

Determination of EST is the first step in the BESIII offline reconstruction software process.
In determining the
time evolution of a track in the MDC, $\est$ is important
for calculating the drift time of the ionization
electrons in a given MDC cell. It is the
basis of the charged track fitting in the MDC~\cite{lab8,lab9}, and its accurate estimation is
essential for further sub-detector reconstruction and
particle identification.
If the $\est$ calculation fails, full tracking\footnote{Full tracking
refers to the charged track fitting algorithm which
exploits the best information in the MDC after the
FST and EST algorithms have been applied.}~\cite{lab8}, will not be implemented due to
the inability to determine the ionization electrons's drift time in any given MDC cell.
So the efficiency of EST
determination is required to be high enough so that the
total detection efficiency is compatible to the
design specification.
Also, incorrect $\est$ may induce
unphysical drift times, which will affect full tracking.
When it is used in data reconstruction, any failure or inefficiency of $\est$ determination
in the EST algorithm also needs to be understood well in MC simulation.
Otherwise, it brings non-negligible systematic uncertainties in the data
analysis. In this paper, a method is introduced to estimate
the efficiency both in data and MC simulation. 

\section{Estimation of efficiencies of determining $\est$}
\subsection{Baseline}
There are two definitions about efficiency of the $\est$ determination which we want to clarify:

\begin{enumerate}
  \item the determination efficiency which is defined as the ratio of events where EST determination successfully returns with available $\est$ information to the total number of events;
  \item the correct determination efficiency which is defined as the ratio of events where EST determination returns correct $\est$ information to the total number of events.
 \end{enumerate}

Studies show that even $\est$ which deviates slightly from
the true value can still be used for full tracking. We therefore take case 1 as the definition of EST efficiency in this paper.
The effect on tracking efficiency due to the wrong $\est$
can then be included in the study of full tracking efficiency.
The definition in case 1 can be formulated as
\begin{equation}
\label{eq1}
\epsilon=\dfrac{N_{\rm{succ}}}{N_{\rm{all}}},
\end{equation}
where $N_{all}$ refers to the number of all events in a given sample,
the $N_{succ}$ refers to the number of events with available $\est$.

In the reconstructed data, since all the selected charged events has successful $\est$,
there is no practical way to select appropriate control sample of the events for the denominator $N_{\rm{all}}$.
In other words, $N_{\rm{all}}$ is dependent on the efficiency of EST determination.

An alternative method must therefore be derived, first to estimate the EST
efficiency of each track in an event and
then to combine the efficiencies of those tracks to estimate
the EST efficiency at the event level.

\subsection{Estimation of EST efficiency at the track level}

The EST efficiency of a track of interest can be studied with the
control samples selected by tagging the other tracks in an event with
the track of interest \emph{missing}. The other tracks serve to tag the event topology with strict requirements to suppress backgrounds. In order to estimate the EST efficiency of the interest track,
all the detector responses from those \emph{tagging} tracks, including
hits in MDC, TOF and EMC, are labeled. $\est$ is then recalculated
with those labeled hits blocked. The EST efficiency of a track
of interest can therefore be estimated by:
\begin{equation}
\label{eq2}
\epsilon^{ \rm{trk}}=\dfrac{N_{(\rm{tag, succ})}}{N_{(\rm{tag})}},
\end{equation}
where $N_{(\rm{tag})}$ stands for the total number of events in the
selected control sample, and $N_{(\rm{tag, succ})}$ for events where
$\est$ is available using the information of the interest track only.
Fig.~\ref{fig1} shows the algorithm flow to estimate
$\epsilon^{\rm{trk}}$, after EST recalculation, events with available
$\est$ are included in $N_{(\rm{tag, succ})}$.

\begin{center}
\includegraphics[width=4cm]{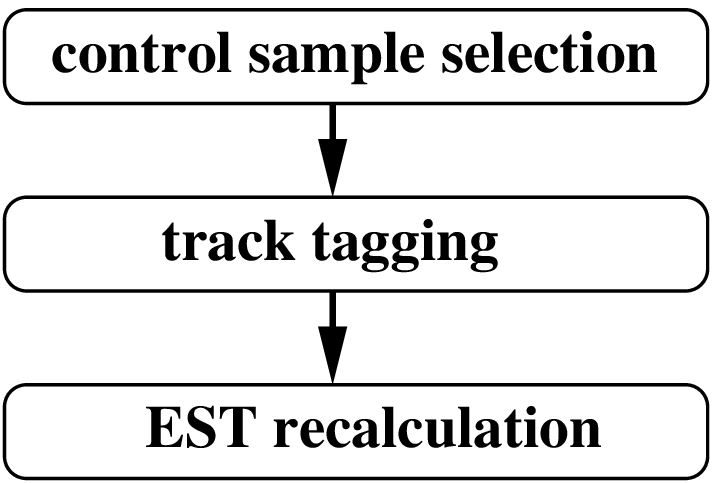}
\figcaption{\label{fig1} Algorithm flow to estimate EST efficiency at the track level.}
\end{center}

One potential effect is that inefficient labeling
of the hits may influences the estimation of $\epsilon^{\rm{trk}}$.
This effect can be studied by evaluating $\epsilon^{\rm{trk}}$ using MC simulation
of single-track events, which are free from the inefficiencies in
labeling. By comparing  $\epsilon^{\rm{trk}}$ from single-track MC
\footnote{The single-track MC is sampled according to the distributions of
transverse momentum or energy and polar angle of the to-be-compared control sample.}
to that from the inclusive MC control sample, the magnitude of
the effect can be understood.

Usually, the EST efficiency, $\epsilon^{\rm trk}$, is shown as a function of
transverse momentum $P_t$ and polar angle $\theta$ for charged tracks. Here, we show $\epsilon^{\rm trk}$ for single $\pi^-$ MC and $\pi^-$ from exclusive MC of $J/\psi\rightarrow \pi^{+}\pi^{-}\pi^{0}$ in Fig.~\ref{fig2}. The bins of $|cos\theta| \approx$ 0.8 correspond the gap of TOF, where TOF information is unavailable, $\est$ could only provided by MDC. That's why the efficiencies at these bins are a little lower. And differences between these points are slightly significant maybe because $\epsilon^{ \rm{trk}}$ are more sensitive to mis-labeling at these bins. However, the overall minor differences indicates that the aforementioned potential effect is negligible.
\end{multicols}
\begin{center}
\includegraphics[width=13cm]{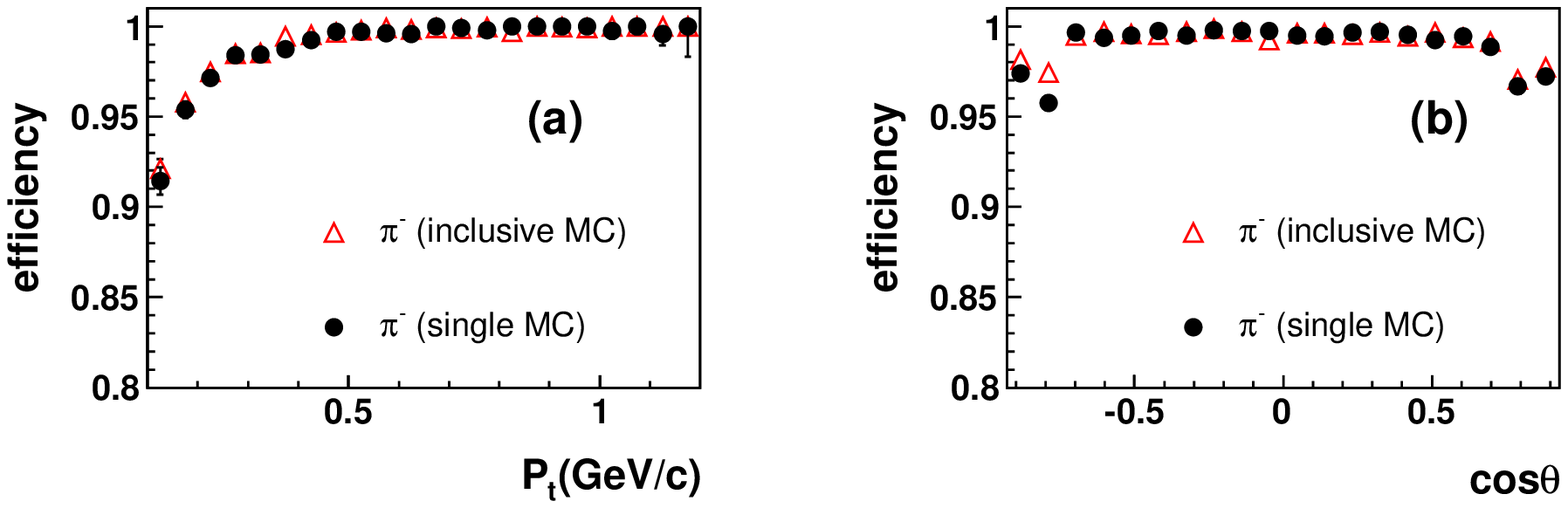}
\figcaption{\label{fig2}EST efficiencies for single $\pi^-$ and $\pi^-$ from inclusive MC as a function of (a) transverse momentum and (b) polar angle.}
\end{center}
\begin{multicols}{2}
\subsection{Estimation of EST efficiency at the event level}

With $\epsilon^{\rm{trk}}$ estimated for control samples of different particle
types, the efficiency at the event level, $\epsilon$, is evaluated
by combining them in a Bayesian way. That is, for a given
physics process, $\epsilon$ of a given event, $i$, can be obtained as follows:
\begin{equation}
\label{eq3}
\epsilon_{i}=1-\prod_{j}(1-\epsilon^{trk}_j),
\end{equation}
where $j$ denotes tracks involved in this event. By averaging efficiencies
over all the exclusively simulated events, we can get the total efficiency
$\epsilon$ for a given process.

Eq.~(\ref{eq3}) is based on the assumption that $\epsilon^{\rm trk}$
for each track is independent. Effects caused by correlations
among tracks of a event can be estimated using an exclusive MC sample by comparing the
efficiency $\epsilon_{\rm{MC}}$, determined with Eq.~(\ref{eq3}),
and the efficiency $\epsilon_{\rm direct}$, which is directly
estimated with Eq.~(\ref{eq1}). An example of such a comparison is
presented in Section 3, showing that the correlation effect is negligible.

\section{Results}
$\epsilon^{\rm trk}$ for charged tracks from data are presented in Fig.~\ref{fig3}.
Also, we use the correction factor $f_{corr}$ to describe the difference of $\epsilon^{trk}$ between data and MC, which is defined as:
\begin{equation}
\label{eq4}
f_{corr}=\frac{\epsilon_{data}^{trk}}{\epsilon_{MC}^{trk}},
\end{equation}
$f_{corr}$ results for charged tracks are presented in Fig.~\ref{fig4}, which shows $\est$ efficiencies from data and MC are
basically consistent for various tracks. For clarity, it is need to point out that for most charged physics process, the contribution from photon to EST determination at the event level is so minor, results for photon are not present in this paper.

\end{multicols}
\ruleup
\begin{center}
\includegraphics[width=7cm]{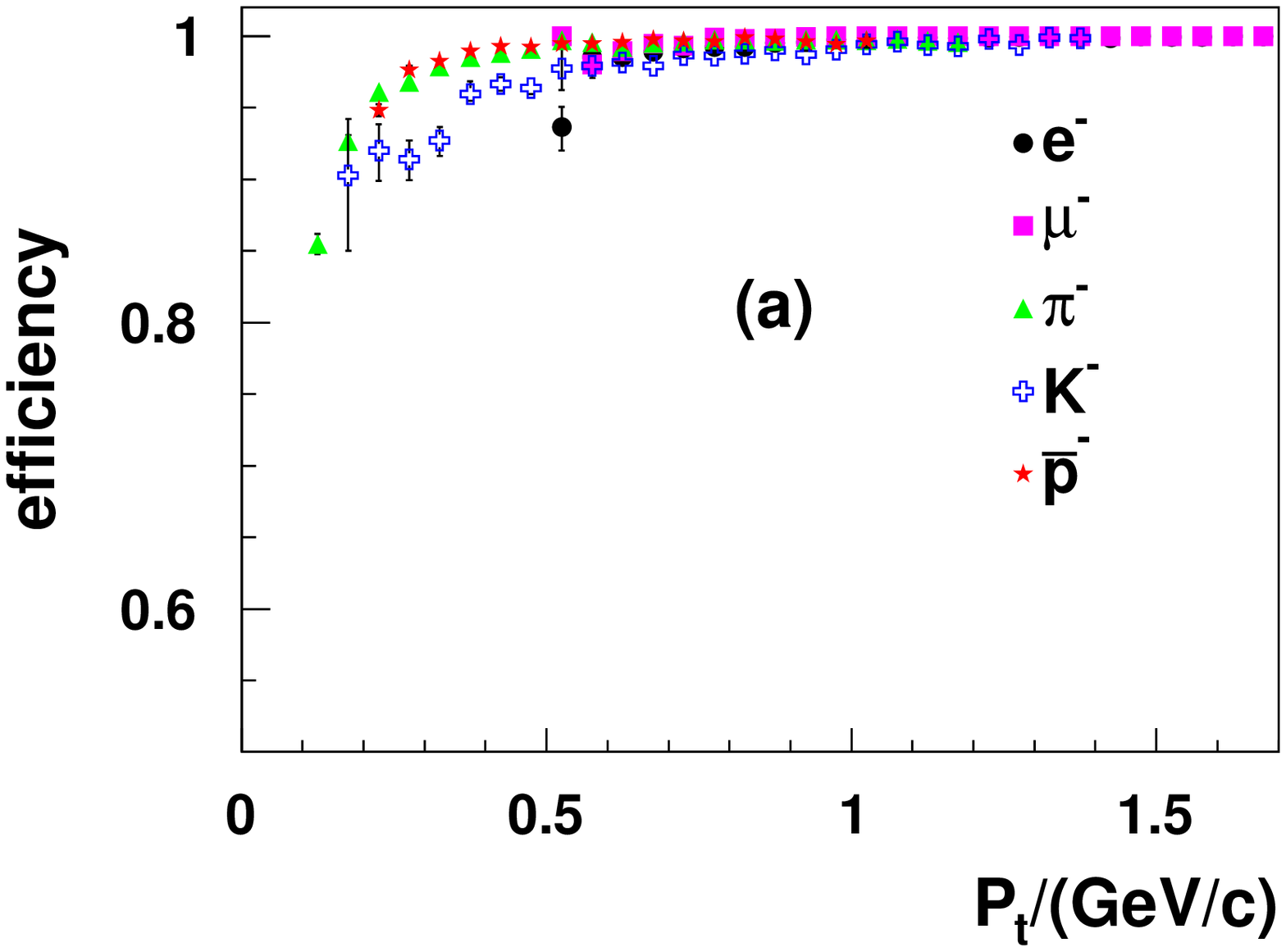}
\includegraphics[width=7cm]{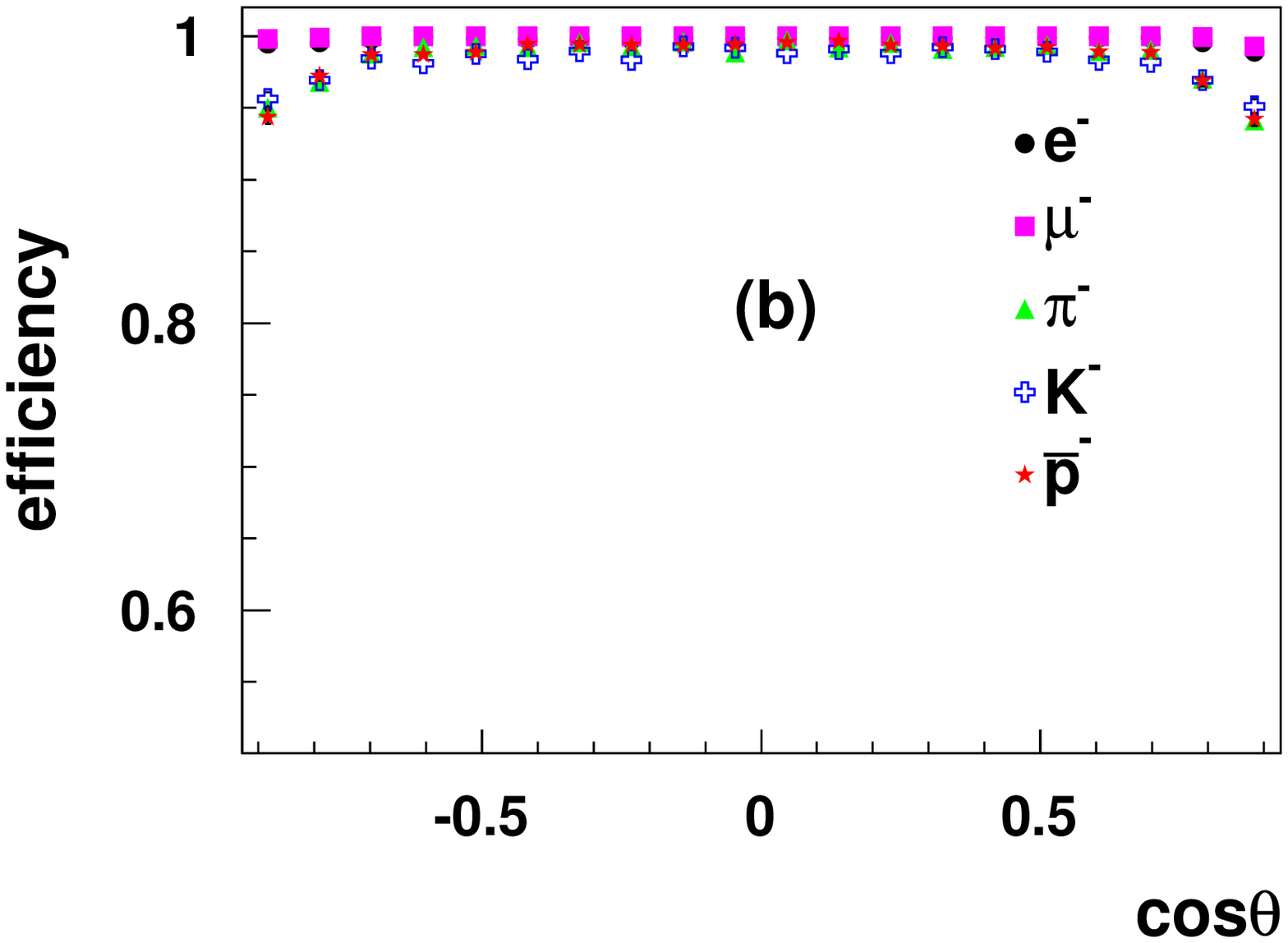}
\figcaption{\label{fig3} EST efficiencies for data at the track level as a function of
 (a) transverse momentum and (b) polar angle.}
\end{center}

\begin{center}
\includegraphics[width=7cm]{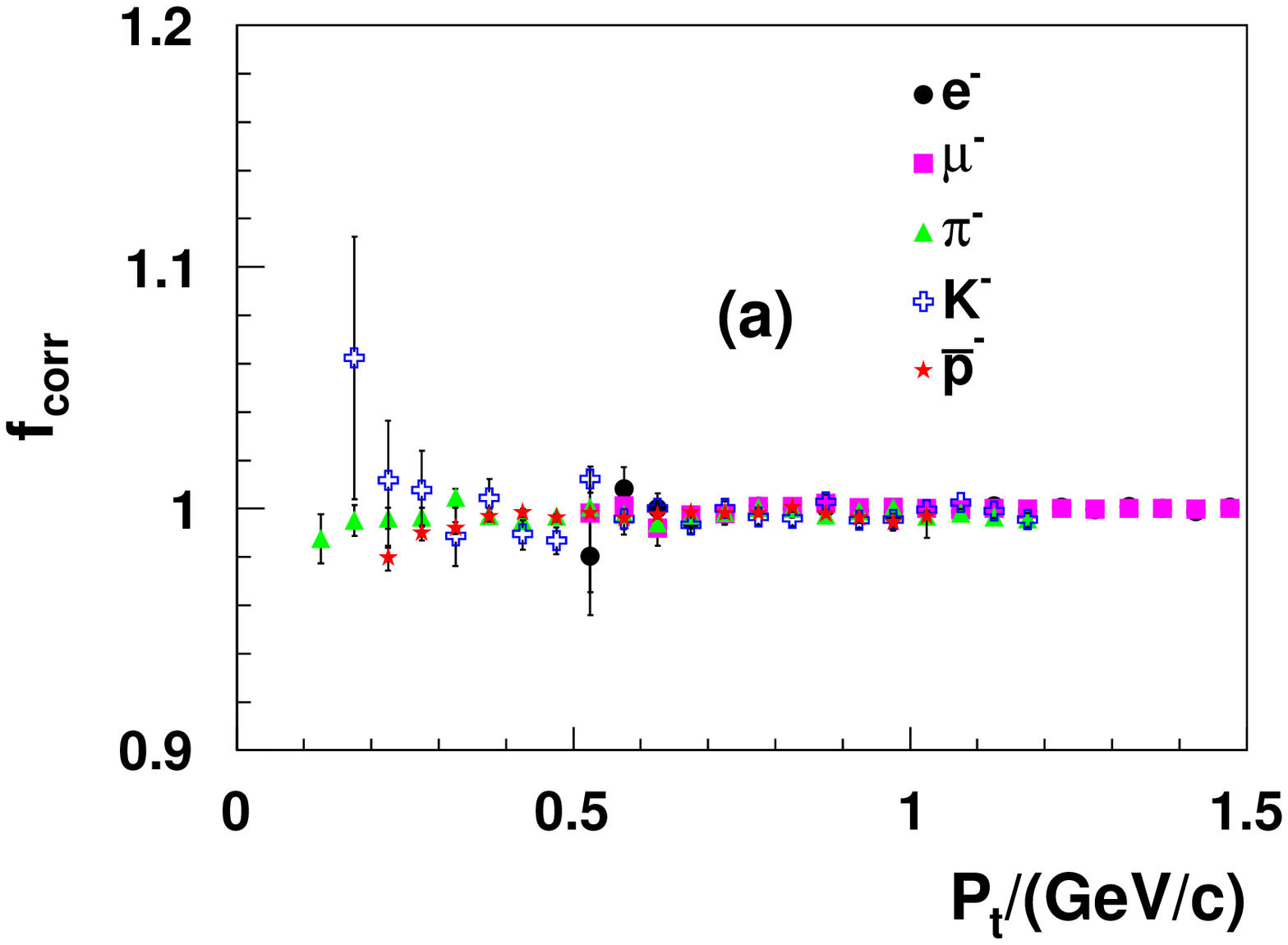}
\includegraphics[width=7cm]{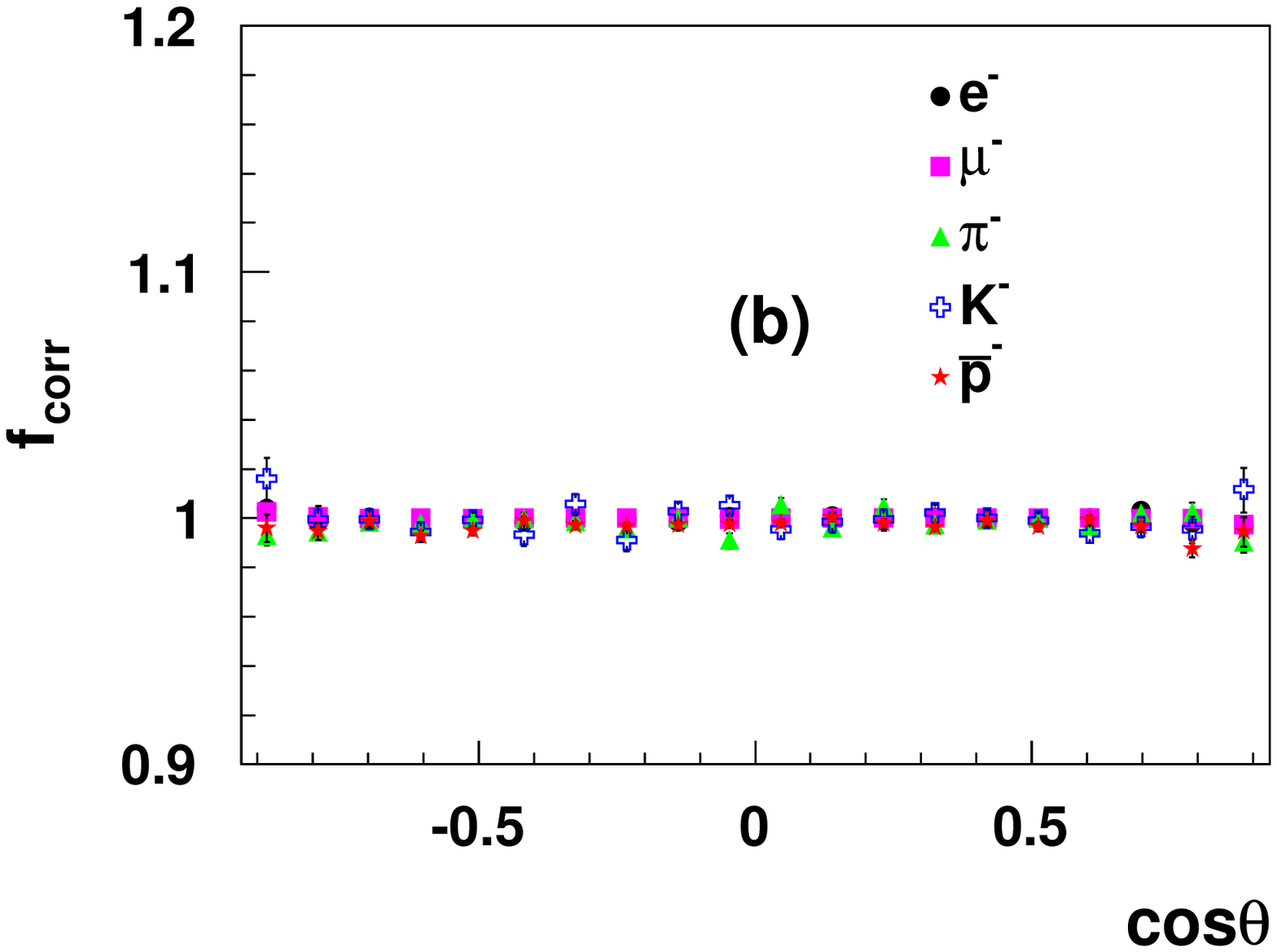}
\figcaption{\label{fig4}The correction factor as a function of (a) transverse momentum and (b) polar angle.}
\end{center}
\ruledown
\begin{multicols}{2}

The EST efficiencies of tracks obtained using this
method can be applied to most physics process. As an example, we take a
typical process which may suffer heavily from
low EST efficiency: the $\psi'\rightarrow\pi^+\pi^- J/\psi$, $J/\psi\rightarrow\gamma\gamma\gamma$ process,
which has two soft pions and three photons. We used 100000 exclusive MC events, as shown in Tab.~\ref{tab1}, the result of $\epsilon_{MC}$ from MC efficiency is consistent with $\epsilon_{direct}$ within statistical errors,
which indicates the negligible correlation effect
among the tracks. Possible systematic uncertainties caused by EST determination is $\epsilon_{data}$/$\epsilon_{MC}-1$, which is found to be about 0.2\% for this process and is negligible~\cite{lab10}.

\begin{center}
\tabcaption{ \label{tab1} Estimation of $\est$ efficiency for $\psi'\rightarrow\pi^+\pi^- J/\psi, J/\psi\rightarrow\gamma\gamma\gamma $. Uncertainties shown are statistical.}
\footnotesize
\begin{tabular*}{80mm}{c@{\extracolsep{\fill}}ccc}
\toprule & $\epsilon_{direct}$ & $\epsilon_{MC}$ & $\epsilon_{data}$\\
\hline
& 98.54$\pm$0.30 (\%)	&  98.72$\pm$0.10(\%) & 98.55$\pm$0.10(\%)\\
$\epsilon_{data}$/$\epsilon_{MC}$  & & \multicolumn{2}{c} {$99.83\pm0.14$(\%)} \\
\bottomrule
\end{tabular*}
\end{center}

\section{Conclusion}

A method to estimate EST efficiency has been established.
For any event topology, its EST efficiency can be
determined by performing a mathematical combination of EST efficiencies for the different tracks in the event.
We present efficiency results for data and correction factors for MC events.
Most of physical process could use these results.
However, there are also two cases which need to be specially treated: events consisting of only
low momentum tracks and photons, and events with tracks
originating from a secondary vertex. In the first case,
the differences in EST efficiencies between real data and MC may be
a little significant; in the second case, the efficiencies
presented in this paper may not be appropriate. Thus, it is
advisable to perform careful studies with the method
demonstrated in this paper if needed.

As discussed before, the effect on tracking efficiency
due to a wrong $\est$ value can be included in the full
tracking efficiency studies. However, events with $\est$ which deviates considerably
from the true event start time could lose all tracks and cannot be included in the full
tracking. Further study may be needed to understand these cases.

\end{multicols}
\vspace{105mm}
\vspace{-1mm}
\centerline{\rule{80mm}{0.1pt}}
\vspace{2mm}

\begin{multicols}{2}

\end{multicols}

\clearpage
%\end{CJK*}
\end{document}